\documentclass[12pt]{article}  
\usepackage{makeidx}
\usepackage{graphicx}
\usepackage{amsmath} 
\usepackage{verbatim}   
\usepackage{rotating} 
\usepackage{fancyvrb}
\usepackage{float}
\usepackage[semicolon, authoryear]{natbib} 
\usepackage{url}  
\usepackage{algorithm}
\usepackage{algpseudocode} 
\usepackage{algorithmicx}
\usepackage{fancyhdr} 
\usepackage{tabularx,booktabs}
\usepackage{tikz}
\usetikzlibrary{calc}

\newcommand{\blind}{01}

\begin{document}

	\def\spacingset#1{\renewcommand{\baselinestretch}%
		{#1}\small\normalsize} \spacingset{1}

	 \title{\bf Distributional Null Hypothesis Testing with the T distribution}
		\author{Fintan Costello \\ School of Computer Science and
			Informatics,\\ University College Dublin\\
			and \\
			Paul Watts \\ Department of Theoretical Physics,\\National University of
			Ireland  Maynooth\\ }
		\date{}
		\maketitle
		\thispagestyle{empty}
\vspace{2cm}

\bigskip

\if1\blind
{	
	\title{\bf Distributional Null Hypothesis Testing }
	\date{}
	\maketitle
	\thispagestyle{empty}
	\vspace{2cm} 
} \fi

\bigskip

\newpage
	
	\bigskip  	
	\begin{abstract} 
	 Null Hypothesis Significance Testing (NHST) has long been central  to the scientific project, guiding theory development and supporting evidence-based intervention and decision-making.   Recent years, however, have seen  growing awareness of serious problems with NHST as it is typically used, and hence to proposals to limit the use of NHST techniques,  to abandon these techniques and move to alternative statistical approaches, or even to ban the use of NHST entirely.  These proposals are premature, because the observed problems with NHST all arise as a consequence of a contingent and in many cases incorrect choice: that of NHST testing against point-form nulls.  We show that testing against distributional, rather than point-form, nulls is better motivated mathematically and experimentally, and that the use of distributional nulls addresses many problems  with the standard point-form NHST approach.  We also show that use of distributional nulls allows a form of null hypothesis testing that takes into account both the statistical significance of a given result and the probability of replication of that result in a new experiment.   Rather than abandoning NHST, we should use the NHST approach in its more general form, with distributional rather than point-form nulls. 
  
	\end{abstract}
 	{\it Keywords:  NHST; Replication; Generalisation }
	\vfill

	\newpage
	\spacingset{1.45} 

\newcommand{\XYlabeledImage}[3]{
	\begin{tikzpicture}
	\node[inner sep=0pt] (A) {#3 };
	\node[black] (B) at ($(A.south)!-0.05!(A.north)$) {#1};
	\node[black,rotate=90] (C) at ($(A.west)!.05!(A.east)$) {#2};
	\end{tikzpicture}}
 
\newcommand{\citeg}[1]{\citeauthor{#1}'s}
\newcommand{\citen}[1]{\citeauthor{#1}}

\begin{quotation}
	{\itshape	In relation to the test of significance, we may say
		that a phenomenon is experimentally demonstrable when we know how to
		conduct an experiment which will rarely fail to give us a statistically
		significant result. \, \citep{fisher1960design}}
	
\end{quotation}

Experimental results are useful when they demonstrate phenomena or effects in such a way as to suggest that these effects are real. This demonstration is especially important in research domains involving complex, interacting, and only partially understood systems (domains such as psychology, medicine, neuroscience, and so on).    Null Hypothesis Significance Testing (NHST) gives a mechanism for assessing the degree to which a given result is inconsistent with a  statistical model, operationalised in terms of the probability $p$ of a result as or more extreme arising under that model.  Taking the null hypothesis of chance variation but no real effect as a statistical model, we can use NHST  to decide whether an observed result suggests a real effect: if this probability $p$ is less than some significance criterion $\alpha$ we take the result as inconsistent with the null model and so as suggesting some effect beyond random variation.

The NHST approach has become a central part of the scientific project, with statistical significance guiding evidence-based intervention and decision-making and   driving theoretical development and understanding.  Despite this, it is becoming increasingly clear that NHST, at least as it is typically used, is fundamentally flawed.  These flaws are apparent when we consider the `replication crisis' in  psychology and related areas (the finding that many statistically significant experimental results are not significant in replications), a crisis seen as `reflecting an unprecedented level of doubt among practitioners about the reliability of research findings in the field' \citep{pashler2012editors}.  These flaws are also evident in well-known effects of sample size on significance (the observation that the probability of getting a statistically significant result in the standard NHST approach increases with sample size, irrespective of the presence or absence of a true effect).  To quote \citet{thompson1998praise}: `Statistical testing becomes a tautological search for enough participants to achieve statistical significance. If we fail to reject, it is only because we've been too lazy to drag in enough participants'.  These flaws are also seen in the repeated observation that the standard NHST approach is overconfident in identifying apparently nonsensical effects (such as telepathy or precognition) as real \citep{wagenmakers2011psychologists}; to quote \citet{diaconis1991replication} `parapsychology is
worth serious study [...because..] it offers a truly alarming massive case study of how statistics can
mislead and be misused'.

In the face of these problems there have been increasing, and increasingly widespread and vocal, calls for the wholesale abandonment of the NHST approach \citep[e.g.][]{BlakeleyAbandon2019,amrhein2018remove,hunter1997needed,carver1978case}.  We think this is premature.  Our aim in this paper is to show that these problems arise from a single source: the use of `point-form' null hypotheses in standard NHST.   We describe an alternative approach to NHST, based on distributional rather than point-form null hypotheses, which addresses the mathematical and experimental problems with point-form NHST, and allows for appropriately conservative estimates of statistical significance.   This distributional NHST approach further allows researchers to give coherent estimates of the probability of replication of statistically significant results, and the degree to which those results will generalise across experiments. 

The structure of this paper is as follows.  We begin by explaining in general terms what we mean by significance testing and replication, and stating some basic assumptions behind the application of distributional NHST. In Section \ref{sec:point_form} we give a brief introduction to significance testing against standard or `point-form' null hypotheses, and explain why these problems with sample size, overconfidence and replication arise with this approach.  In Section \ref{sec:distributional_form} we present the distributional NHST approach, and show how this approach addresses these problems.  The distributional null approach involves the estimation of cross-experiment variance; in Section \ref{sec:q_estimation} we estimate  this variance from experimental data.  In section \ref{sec:joint_criterion} we discuss approaches to applying the distributional null hypothesis testing in practice, and in the final section we address possible criticisms of this  distributional null approach.

\section{Terminology and assumptions}
\label{sec:assumptions}

Ideas of NHST have been interpreted and used in many different ways in the literature.  We take a specific interpretation: we take a `null hypothesis significance test' to be a test that assesses the probability of a given experimental result, or a more extreme result, arising purely as a consequence of random variation in a statistical model that assumes no effect.     Our focus here is solely on simple experiments involving comparison of means; that is, one-sample, paired, or two-sample experimental designs. NHST in these experiments involves calculating the probability of obtaining a value as or more extreme than the observed mean $x$ under a particular null model.  If this probability is less than some criterion $\alpha$ then the researcher can conclude that the result may not be solely a consequence of chance variation as described in that model:  the experimental result is significant, relative to that model, because it suggests the presence of some non-random `real' effect.  

  What do we mean by `real' here?  We could take `real' to mean `unlikely to be a consequence of random variation'.  This is not quite satisfactory, however: a `real' result should be one that will be reliably demonstrated in repeated experiments, and the fact that a given result is unlikely to be a consequence of random variation does not, in itself, tell us anything about replication.   Ideally we would like to be able to say that a given experimental result is `real'  in terms of  statistical significance (a result is real because it is unlikely to be due to chance), \textit{and} in terms of replication (a result is real because it is experimentally demonstrable:  similar results will occur reliably in repeated experiments).   To express this concretely: we would like to count an experimental result as real when $p < \alpha$ and where the probability of obtaining the same result, also with $p < \alpha$, in a replication of this experiment is greater than some criterion $\beta$.
 
 What do we mean by `the same result'?   Where statistical significance by itself can be two-sided (a result may be statistically significant if it is greater than expected under random variation, or if it is less than expected),  replication  can only be one-sided: if a result $x_1$ is statistically significant and less than expected under chance, and a result $x_2$ is statistically significant but greater than expected under chance, then $x_2$ is not the same result as $x_1$ and does not count as a replication of that result.   We thus take two experimental results $x_1$ and $x_2$ to be the same  when both results achieve statistical significance at the same level $\alpha$ and when both results go in the same direction: either both are less than expected under the null hypothesis, or both are greater than expected.  Note that in this view, the question of replication is meaningless for  `omnibus' or `global' statistical tests of variance (such as the F-test), because such tests do not consider the direction of deviation from  the null hypothesis (variance being always positive, whether the direction of the difference is positive or negative). 
 
What do we mean by `repeated experiments'?   We assume that a repeated experiment is one which matches the original exactly in terms of the variable $X$ being measured (both original and repeat are attempting to measure exactly the same variable), experimental design (both use the same design), and sample size.  Beyond this, experiments can vary in many different ways, differing in characteristics of the source population, in the specific materials used, in the precise measurement procedure employed, and so on.  We would like to specify the degree to which a statistically significant result should be expected to replicate across variation in these factors; that is, to generalise to experiments that vary, to a greater or lesser extent, from the original.  To express this concretely: we would like to count an experimental result as real when that result is statistically significant at criterion $\alpha$, where the probability of replication of that result that is greater than $\beta$, under the assumption that cross-experiment variance is within some generalisation criterion $\gamma$.
  
 Finally, note that questions about the probability of replication of a given experimental result are only really of interest when the result is new.  If a given result has been investigated in a number of previous experiments and has reached statistical significance in some proportion $R$ of those experiments, we already implicitly know that result's probability of replication is close to $R$.  Given this, our discussion of significance, replication, and generalisation focuses on experimental results that are new; that are not themselves replications of previous studies.

 \section{Standard or `point-form' null hypothesis testing}
 \label{sec:point_form}

 We begin by briefly illustrating problems with standard or point-form null hypothesis testing, using the $t$ distribution. 
 
  We consider a situation where experimental results consist of $N$ measurements which we assume follow some normal distribution $X \sim \mathcal{N}(\mu,\,\sigma^2)$ with unknown variance $\sigma^2$ (the situation of a one-sample $t$ test). The distribution of the sample mean of this variable is 
\begin{equation*}
\overline{X} \sim \mathcal{N}\left(\mu,\,\sigma^2/N\right)
\end{equation*}
and so the variable  
\begin{equation*}
\frac{\overline{X}}{(\sigma/\sqrt{N})}
\end{equation*}
is normally distributed with mean $0$ and variance $1$.  Let the variable
\begin{equation*}
S^2= \frac{1}{N-1}\sum_{i=1}^{N}(X_i-\overline{X})^2
\end{equation*}
represent the sample variance of $X$, and the variable 
\begin{equation*}
\frac {(N-1)\ S^{2}}{\sigma^{2}}
\end{equation*}  
has a chi-squared distribution with $\nu = N - 1$ degrees of freedom.  Let $Z = \overline{X}/S$ 
represent the ratio of sample mean to sample standard deviation (the normalised sample mean), and we see that the variable
\begin{equation*} 
T=  \frac{ \frac{\overline{X}}{(\sigma/\sqrt{N})}}{\sqrt{\frac{(N-1)\ S^{2}}{\sigma^{2}/(N-1)}}}  =Z\sqrt{N} 
\end{equation*}  
has a $t$ distribution with $\nu=N-1$ degrees of freedom.
 
We assume that the point-form null hypothesis is that $\mu$ has the value $\mu=0$.  Letting $z = \overline{x}/s$ 
be the ratio of sample mean to sample standard deviation obtained in a given experiment,  the $p$-value of this result $z$ relative to the point-form null (that is, the probability of obtaining a result as or more extreme than $z$) is
\begin{equation*} 
1- T_{\nu}\left( |z| \sqrt{N} \right)
\end{equation*}
 where $T_{\nu}$ is the cumulative probability of the $t$ distribution with $\nu$ degrees of freedom. For a given significance level $\alpha$  we define the critical value $Z_{crit}$ as the value such that
\begin{equation*}
   T_{\nu}(Z_{crit} \sqrt{N}) = 1-\alpha
\end{equation*}
or equivalently
\begin{equation*}
Z_{crit}  = T_{\nu}^{-1}(1-\alpha)/\sqrt{N}
\end{equation*}
and a observed value  $z$ will be statistically significant at level $\alpha$ relative to the point-form null when $|z| \geq Z_{crit}$.

As $N$ increases this bound $Z_{crit}$ approaches $0$ and so, for large enough $N$,  any $z \ne 0$ will be counted as statistically significant.   However, the probability of $z$ being exactly equal to $0$ is vanishingly small: the value of $z$ will vary around $0$ to some degree, even if there is no real effect.  This means that the probability of getting a statistically significant result under a point-form null hypothesis is  an increasing function of the sample size $N$. 
 To quote \citet{cohen2016earth}
\begin{quote}``[the point-form null hypothesis ] can only be true in the bowels of a computer processor running a Monte Carlo study (and even then a stray electron may make it false). If it is false even to a tiny degree, it must be the case that a large enough sample will produce a significant result and lead to its rejection. So if the null hypothesis is always false what's the big deal about rejecting it?"
\end{quote}
This problem undermines point-form null hypothesis testing as a method for identifying potentially real effects (with large enough sample size, any experiment is likely to produce a `real' effect) and so explains the relatively low levels of replication for statistically significant results \citep[e.g.][]{camerer2018evaluating,open2015estimating,klein2018many,klein2014investigating}. This problem also explains the observation of statistically significant results in parapsychological studies, which typically involve very large sample sizes \citep[and where significance is often inversely related to sample size; see e.g. ][]{bosch2006examining}.  


A separate problem concerns the question of replication of significant results. Assume that we have observed a statistically significant result in our first experiment, and wish to estimate the probability that a repeat of that experiment (with the same sample size) will also give a statistically significant result at the same level $\alpha$. The standard approach to estimating this probability of replication under the point-form null hypothesis is via statistical power, so that the probability of replication of a result $t_1$ in an exact replication with degrees of freedom $\nu$ is approximately
\begin{equation*}
1- \Phi\left(\frac{T_{\nu}^{-1}(\alpha)-t_1}{\sqrt{1 + \frac{T_{\nu}^{-1}(\alpha)^2}{2\nu}}}\right)
\end{equation*}

 \citep[with the cumulative normal being used to approximate the non-central $t$ distribution; see e.g.][]{greenwald1996effect,posavac2002using,gorroochurn2007non}.   This approach assumes that, since a significant result has been obtained in experiment $1$,  we should conclude that the null hypothesis $\mu = 0$ is  false and the alternative hypothesis  $\mu = \overline{x} $ is  true.  Replication is then measured in terms of the probability of getting a statistically significant result relative to the null hypothesis $\mu = 0$, under the assumption that, in fact, $\mu = \overline{x}$. 

The difficulty here is that, even with a statistically significant result, we cannot conclude with any confidence that the specific point-form hypothesis $\mu = \overline{x}$ is, in fact, true.  There is still some probability that $\mu = 0$; and indeed for every possible value, across the entire range, there is some probability that $\mu$ is equal to that value (with this probability increasing as values approach $\overline{x }$).  Our estimate of the probability of replication of a significant result should be based on this probability distribution for $\mu$, whatever it is.  Such a probability distribution for $\mu$ cannot be derived from point-form  hypotheses, which do not assign any probability to values $\mu \neq 0$ (for the null hypothesis) or  $\mu \neq \overline{x}$ (for the alternative hypothesis), and so it is not clear whether the above power expression gives a correct estimate of the probability of replication of a given result. 

These problems are mathematical in form.  An final problem with the use of point-form nulls arises from the practical problem of experimental design.    One central task for an experimenter investigating a particular effect is the control of confounding factors: factors which are of no theoretical interest, but which may influence the experimental mean $\mu$ for a given experiment in some way.    These confounding factors can never be completely controlled and so, even if there is truly no effect, the experimental mean $\mu$ will vary randomly across experiments due to variation in the influence of these factors.  The point form null hypothesis, however, assumes that $\mu=0$: that such confounds are perfectly controlled.  This is an unrealistic picture of the experimental process especially in areas investigating complex, interacting, and only partially understood systems, where results are necessarily subject to many difficult-to-control confounds.  
 
The above problems all arise because the point-form null hypothesis is  flawed as a statistical model of random variation in experimental results.   Under a point-form null, the statistical model is that all experiments have exactly the same mean $\mu=0$, and any difference between an observed  $\overline{x}$ and $0$ in a given experiment arises solely a consequence of random variation in sampling \textit{within that experiment}.   This model ignores the possibility that a difference  between  observed mean $\overline{x}$ and expected value $0$ in a given experiment may arise, in the absence of any real effect, simply because of random variation in $\mu$ across experiments.  This point-form model, in other words, systematically underestimates the degree of random variation affecting experimental results.   In the next section we describe an alternative `distributional' null hypothesis model which addresses these issues by assuming random variation in means both within and across experiments. 

\section{Testing against a distributional null hypothesis}
 \label{sec:distributional_form}
 
As before, we consider experimental results consisting of $N$ measurements which we assume follow some normal distribution $X \sim \mathcal{N}(\mu,\,\sigma^2)$ with unknown variance $\sigma^2$ (the situation of a one-sample $t$ test).   We also assume that values of $\mu$ across experiments themselves vary randomly and are drawn from a normal distribution $\mu \sim \mathcal{N}\left(\mu_0,\, \sigma_0^2\right)$.
  We characterise the degree of cross-experiment variance $\sigma_0^2$   via the variance ratio 
  \begin{equation*}
  q = \frac{\sigma_0^2}{\sigma^2} 
  \end{equation*}
  Our general null hypothesis is that $\mu_0=0$, and so a distributional null hypothesis is that values of $\mu$ across experiments are distributed as $\mu \sim \mathcal{N}\left(0,\, q \,\sigma^2\right)$:  a specific choice of value for $q$ represents a specific assumption about the relationship between cross-experiment and within-experiment variance.     

 For a given distributional null (for a given value of $q$) we have 
  \begin{equation*}
  \begin{split}
  p(\overline{X}|N,\sigma, q) &=  \int \mathcal{N}(\overline{x}| \mu,\,\sigma^2/N) \mathcal{N}\left(\mu | 0,  \sigma_0^2\right) d\mu =    \mathcal{N}\left(\overline{X}| 0,\, \frac{\sigma^2}{N}+ q\sigma^2\right) \\ 
  &=    \mathcal{N}\left(\overline{X}| 0,\, (\sigma^2/N)(1+ qN) \right) \\  
  \end{split}
  \end{equation*}
   \citep[a property of the Normal distribution; see e.g.][]{bishop2006pattern,murphy2007}.
   This means that the variable 
   \begin{equation*}
   \frac{\overline{X}}{(\sigma/\sqrt{N})\sqrt{1+qN}}
   \end{equation*}
   is normally distributed with mean $0$ and variance $1$.  As before, let the variable
   \begin{equation*}
   S^2= \frac{1}{N-1}\sum_{i=1}^{N}(X_i-\overline{X})^2
   \end{equation*}
   represent the sample variance of $X$ within a given experiment, and the variable 
   \begin{equation*}
   \frac {(N-1)\ S^{2}}{\sigma^{2}}
   \end{equation*}  
   has a chi-squared distribution with $\nu = N - 1$ degrees of freedom.  Letting $ Z = \overline{X}/S$ 
   represent the normalised sample mean, and we see that the variable
   \begin{equation*} 
   T=  \frac{ \frac{\overline{X}}{(\sigma/\sqrt{N})\sqrt{1+qN}}}{\sqrt{\frac{(N-1)\ S^{2}}{\sigma^{2}/(N-1)}}}  =\frac{Z\sqrt{N}}{\sqrt{1+qN}} 
   \end{equation*}  
   has a $t$ distribution with $\nu=N-1$ degrees of freedom.   With $z = \overline{x}/s$ being the ratio of sample mean to sample standard deviation obtained in a given experiment, the $p$-value of  $z$ relative to the distributional null $q$ is
   \begin{equation}
   \label{eq:distributional_p}
   1-T_{\nu}\left( \frac{ |z| \sqrt{N}}{\sqrt{1+qN}}  \right)
   \end{equation}
   For a given significance level $\alpha$  we define the critical value  $Z_{crit}$, the significance bound relative to the distributional null $q$, as the value such that
   \begin{equation*}
   T_{\nu}\left( \frac{ Z_{crit} \sqrt{N}}{\sqrt{1+qN}}\right) = 1-\alpha
   \end{equation*}
   and so
   \begin{equation} \label{eq:Z_crit_q}
   Z_{crit}  = \frac{T_{\nu}^{-1}(1-\alpha)\sqrt{1+qN}}{\sqrt{N}}
   \end{equation}
   and a observed value  $z$ will be statistically significant at level $\alpha$ relative to the distributional null $q$ when $|z| \geq Z_{crit}$.      Notice that from Equation \ref{eq:Z_crit_q} we have the bound 
  \begin{equation} \label{eq:X_crit_bound}
  Z_{crit} > T_{\nu}^{-1}(1-\alpha)\sqrt{q}
  \end{equation}
  for all $N$; as sample size $N$ increases, $Z_{crit}$ approaches this bound.   Values of $|z|$ less than this bound will never reach significance relative to distributional null $q$, no matter how large the sample size $N$: this bound represents a rejection region for the null hypothesis, which is constant for fixed $q$ (but falls monotonically as $q$ declines).    This distributional significance test thus does  not suffer from the problems associated with sample size that arise with point-form significance testing.  Putting this another way, under a distributional null $q$ there is a direct link between statistical significance and effect size: only effect sizes greater than  this bound can ever achieve significance.

  \subsection{Replication in distributional NHST}
  
  Up to this point we've expressed experimental results in a quasi-normal form $Z$, representing the sample mean in units of sample standard deviation.  At this point we convert to the standard $t$ form, so that
  \begin{equation*}
  t_1 = \frac{\overline{x}}{(s/\sqrt{N})} = z_1 \sqrt{N}
  \end{equation*}  
  and we have the significance criterion for a test with degrees of freedom $\nu$ of
  \begin{equation} \label{eq:T_crit_q}
  t_{crit}  = T_{\nu}^{-1}(1-\alpha)\sqrt{1+qN}
  \end{equation}
  and the $p$ value of a result $t_1$ relative to the distributional null $q$ is 
  \begin{equation*} 
 1- T_{\nu}\left( \frac{|t_1|}{\sqrt{1+qN}}  \right)
  \end{equation*}
      
Now we consider the probability of replication under this distributional null model, relative to some distributional null $q$.  This is the probability that we will get a significant  result 
in a replication  (a result where $|t_2| \geq t_{crit}$ and $t_2$ and $t_1$ have the same sign),  given our initial result $t_1$.  Our statistical model here is that both sample size $N$ and within-experiment variance $\sigma^2$ are the same in these two experiments, and that we are testing relative to the same null hypothesis $q$ and the same significance level $\alpha$ (and so the same critical value $t_{crit}$) .
 
 Let the variable $\overline{X_1}$ represent the observed sample mean in experiment 1; then conditional on that variable we update our initial distribution for $\mu$ based on this observed data  (the initial distribution being $\mu \sim  \mathcal{N}( 0,\,\sigma_0^2)$).   Standard Gaussian updating \citep[see e.g.][]{murphy2007} gives
 \begin{equation*}
  \mu \sim \mathcal{N}(  \mu_N,\,\sigma_N^2)  
 \end{equation*}
 where
 \begin{equation*}
 \sigma_N^2 = \frac{\sigma_0^2\,  \sigma^2 }{N\, \sigma_0^2 + \sigma^2}  =  \frac{qN}{(1+qN)}  ( \sigma^2/N)   
 \end{equation*}
 \begin{equation*}
 \mu_N =  \frac{\sigma_N^2  }{( \sigma^2/N)}\, \overline{X_1}=  \frac{qN}{(1+qN)} \overline{X_1} 
 \end{equation*}
Given this updated distribution for $\mu$, we can  express the probability density of the sample mean $\overline{X_2}$ in our second experiment as
  \begin{equation*}
  p(\overline{X_2}|\overline{X_1},  q) =  \int \mathcal{N}(\overline{X_2}| \mu,\,\sigma^2/N) \mathcal{N}(\mu | \mu_N,\,\sigma_N^2) d\mu
  \end{equation*}
  and so 
   \begin{equation*}
   \begin{split}
   \overline{X_2} & \sim \mathcal{N}( \mu_N,\,\sigma^2/N+\sigma_N^2) \\
     & \sim \mathcal{N}\left( \frac{qN}{(1+qN)}\, \overline{x_1},\,\frac{\sigma^2(1+ 2qN)}{N(1+qN)}\right) \\
   \end{split}
   \end{equation*}
  and the variable
 \begin{equation*}
  \frac{\overline{X_2}-\frac{qN}{1+qN}\, \overline{X_1}}{\frac{\sigma}{\sqrt{N}}\sqrt{\frac{1+2qN}{1+qN}}} 
 \end{equation*}
 is normally distributed with mean $0$ and variance $1$.  Let $S_2^2$ represent the sample variance in the second experiment
 and  the variable
    \begin{equation*}
 t_2 = \frac{\overline{X_2}}{(S_2/\sqrt{N})}   
 \end{equation*}
 represent the $t$ value for that experiment's result. Since both sample size $N$ and within-experiment variance $\sigma$ are assumed to be the same in both experiment $1$ and experiment $2$, we see that $S$ and $S_2$ follow the same distribution, and so  
     \begin{equation*}
 t_2 \sim \frac{\overline{X_2}}{(S/\sqrt{N})} 
 \end{equation*}  
and so the variable 
   \begin{equation*} 
   \begin{split}
   \frac{\overline{X_2}-\frac{qN}{1+qN}\, \overline{X_1}}{\sqrt{\frac {(N-1)\ S^{2}}{\sigma^{2}}/(N-1)}\frac{\sigma}{\sqrt{N}}\sqrt{\frac{1+2qN}{1+qN}}}  = \frac{t_2-\frac{qN}{1+qN }t_1}{ \sqrt{\frac{1+2qN}{1+qN}}}  \\
\end{split}
   \end{equation*}  
   has a $t$ distribution with $\nu=N-1$ degrees of freedom.  This means that  the probability of getting a result more extreme than our significance criterion $t_{crit}$ in experiment 2, and with the same sign as our original result $t_1$, is 
   \begin{equation} \label{eq:p_r}
   \begin{split}
  p_r(t_1,\nu)  &= 1-T_{\nu}\left(\frac{t_{crit}-\frac{qN}{1+qN }|t_1|}{ \sqrt{\frac{1+2qN}{1+qN}}} \right) \\
  &= T_{\nu}\left(\frac{\frac{qN}{1+qN }|t_1|-t_{crit}}{ \sqrt{\frac{1+2qN}{1+qN}}} \right) 
   \end{split}
   \end{equation}
   
   Equation \ref{eq:T_crit_q} gives a criterion for significance  based on a distribution for $\mu$, while Equation \ref{eq:p_r} gives a probability of replication based on that distribution updated on the value of the observed $\overline{x_1}$.  These expressions thus address the problems with point-form significance testing described above.  In terms of significance, this distributional criterion $t_{crit}$ ensures that the null hypothesis is not `always true'; that significance is not simply a function of sample size, and that negligible effects are not counted as statistically significant.   In terms of replication, the expression $p_r$ assigns a probability density to every possible value of $\mu$ and so avoids the problem with point-form replication estimates based on statistical power (which unrealistically assign all probability density to $\mu = \overline{x_1}$).

We've derived these significance and replication results under the assumption of a single sample of normally distributed data (the assumptions of a one-sample $t$ test).  They extend naturally, however, to paired and independent two-sample $t$ tests with equal sample sizes.

For a paired $t$ test, we assume an experiment involving  $N$ pairs of measurements $x$ and $y$, each drawn from some normal distribution $\mathcal{N}( \mu_X,\sigma^2)$ and $\mathcal{N}( \mu_Y,\sigma^2)$ (with equal variance in $x$ and $y$) and where values of $\mu_X$ and $\mu_X$ vary across experiments and are themselves drawn from the normal distribution $ \mathcal{N}(0,\,q\sigma^2)$ (this is our distributional null hypothesis, as before).  Let variable $d_i$ represent the $x_i-y_i$ difference in each pair.  Since means and variances of sums of normal distributions are additive, values of $d$ are drawn from the normal distribution  $\mathcal{N}( \mu_D,2\sigma^2)$ where $\mu_D$ is itself drawn from the normal distribution  $ \mathcal{N}(0,\,q2\sigma^2)$, as before (the distributional null hypothesis).  Let the variable $\overline{d}$ represent the mean difference and we have
\begin{equation*} 
\overline{d}  \sim \mathcal{N}\left(  0,\,(2\sigma^2/N)(1+ qN)\right)
\end{equation*}
and the variable
\begin{equation*}
Z = \frac{\overline{d}}{\sqrt{(2\sigma^2/N)(1+ qN)}}
\end{equation*}
follows a normal distribution with mean $0$ and variance $1$.
Let 
\begin{equation*}
s_d^2 = \frac{1}{N-1}\sum_{i=1}^{N}(d_i-\overline{d})^2
\end{equation*}
and in a given experiment (with a fixed value of $\mu_D$) the variable 
\begin{equation*}
V=\frac {(N-1)\ s_d^{2}}{2\sigma^{2}}
\end{equation*}  
has a chi-squared distribution with $\nu = N - 1$ degrees of freedom.  Letting
\begin{equation*}
t_1 =   \frac{\overline{d}}{s_d/\sqrt {N}}
\end{equation*} 
we see that the variable
\begin{equation*} 
\frac{Z}{\sqrt{V/\nu }} = \frac{\frac{\overline{d}}{\sqrt{(2\sigma^2/N)(1+ qN)}}}{ \sqrt{\frac {(N-1)\ s_d^{2}}{2\sigma^{2}}/(N - 1)}}=    \frac{t_1}{\sqrt{1+qN}}
\end{equation*}  
has a $t$ distribution with $\nu=N-1$ degrees of freedom.  From this point our derivation runs just as before, giving a significance criterion for $t_1$ as in Equation \ref{eq:T_crit_q} and a probability of replication as in Equation \ref{eq:p_r}. 

Similarly, for an independent two-sample $t$ tests with equal sample size, we assume an experiment involving two independent sets of measurements $X_1$ and $Y_1$, with $N$ measurements in the first set and the same number $N$ in the second, and with each drawn from some normal distribution $\mathcal{N}( \mu_Y,\sigma^2)$ and $\mathcal{N}( \mu_Y,\sigma^2)$ (again, we assume equal within-sample variance $\sigma^2$ in both samples).   Let $\overline{x_1}$ and $\overline{y_1}$ represent the means of these measurements, and in a single experiment (with fixed values of $\mu_X$ and $\mu_Y$) we have
$\overline{x_1}   \sim \mathcal{N}\left(  \mu_X,\,\sigma^2/N\right)$ and  $\overline{y_1}   \sim \mathcal{N}\left(  \mu_Y,\,\sigma^2/N\right)$.  Let $\overline{d_1} $ be the difference between these mean measurements (again in a single experiment) and we have $\overline{d_1}   \sim \mathcal{N}\left(  \mu_X-\mu_Y,\,2\sigma^2/N\right)$.

We assume that values of $\mu_X$ and $\mu_Y$ vary across experiments and are themselves drawn from the normal distribution $ \mathcal{N}(0,\,q\sigma^2)$ (the distributional null), so that, across experiments, the difference $\overline{d_1}  $ follows the distribution
\begin{equation*} 
\overline{d_1}  \sim \mathcal{N}\left(  0,\,2(\sigma^2/N)(1+ qN)\right)
\end{equation*} 
and the variable
\begin{equation*}
Z = \frac{\overline{d_1} }{\sqrt{(2\sigma^2/N)(1+ qN)}}
\end{equation*}
follows a normal deviation with mean $0$ and variance $1$.
Letting 
\begin{equation*}
s_p^2 = \frac{ S_{X_1}^2 +  S_{Y_1}^2 }{2}
\end{equation*}
be the `pooled sample variance' calculated for the current experiment, and the variable
\begin{equation*}
V=\frac {(2N-2)\ s_p^{2}}{\sigma^{2}}
\end{equation*}  
has a chi-squared distribution with $\nu = 2N - 2$ degrees of freedom.  Letting
\begin{equation*}
t_1 =   \frac{\overline{x_1} - \overline{y_1}}{s_p \sqrt {2/N}}
\end{equation*} 
be the $t$ value calculated for an independent two-sample $t$ test with equal sample size under the point-form null hypothesis, and  we see that the variable
\begin{equation*} 
\frac{Z}{\sqrt{V/\nu }} = \frac{\frac{\overline{x_1} - \overline{y_1} }{(\sqrt{(2\sigma^2/N)(1+ qN)}}}{ \sqrt{\frac {(2N-2)\ s_p^{2}}{\sigma^{2}}/(2N - 2)}}=    \frac{t_1}{\sqrt{1+qN}}
\end{equation*}  
has a $t$ distribution with $\nu=2N-2$ degrees of freedom.   From this point our derivation runs just as before, giving a significance criterion for $t_1$ as in Equation \ref{eq:T_crit_q} and a probability of replication as in Equation \ref{eq:p_r}, but with degrees of freedom $\nu = 2N-2$.

\section{Estimating variance ratio $q$}
\label{sec:q_estimation}

For these results to be useful, we need to have some estimate of reasonable values for the variance ratio $q$ that might hold in experimental tasks.  In this section we estimate the distribution of $q$, using data from the first `Many Labs' replication project \citep{klein2014investigating}\footnote{Data available at \url{https://osf.io/wx7ck/}} .

The first `Many Labs' replication project \citep{klein2014investigating} involved the replication of 13 classic and contemporary psychological effects across
36 different sites (36 distinct samples and settings).  We chose to use this dataset to estimate the distribution of the variance ratio $q$ because of these $13$ experiments, $9$ involved participants giving continuous responses of some form (subsequent Many Labs replication projects tended to involve a higher proportion of experiments involving binary or categorical choice measures, to which our analysis does not apply).   For each of these $9$ experiments we identified each such response measure and calculated, for each of $36$ different sites (different labs), the variance ratio $q$ associated with that lab (cross-lab variance in the measure in question, divided by within-lab variance in that measure). 

Of the $36$ different sites involved in this replication study, 25 were based in the US, and 11 were international (occurring Malasyia, Turkey, Italy, Czechia, Poland, Brazil, Canada,and the UK).   Some of the experiments involved materials that were in some way culturally specific in some way (e.g., estimating distances between US cities, estimating political attitudes of typical Americans, investigating priming effects of the American flag).  Our expectation was that for these experiments there would be differences in variance ratio $q$ between results from US and international sites.  

\subsection{Results}

Across these experiments we identified 60 distinct measures, covering topics ranging from estimates of the time of day a photo was taken, to the influence of sunk costs of decision making, to implicit and implicit attitudes towards art and math, estimates of the distance from San Francisco to New York City, the population of Chicago, or the number of babies born in the US per day,  to the level of conservatism of the typical American,  attitudes towards issues such as abortion and gun control, and the degree of justice of the societal system in each site country (details of these measures are given in Appendix A).  For each individual measure  we calculated the variance of that measure within each replication site, and variance in the mean value of that measure across all sites; from this we calculated that measure's variance ratio for each site (overall between-site variance for that measure, divided by that measures variance in the individual site).  We expected these variance ratios to be greater than 0, but relatively low and consistent (since each site was aiming to carry out an exact replication of the given experiment, and so cross-site variance would be expected to be small).   

\begin{figure}
 	\centerline{\scalebox{0.6}{\includegraphics{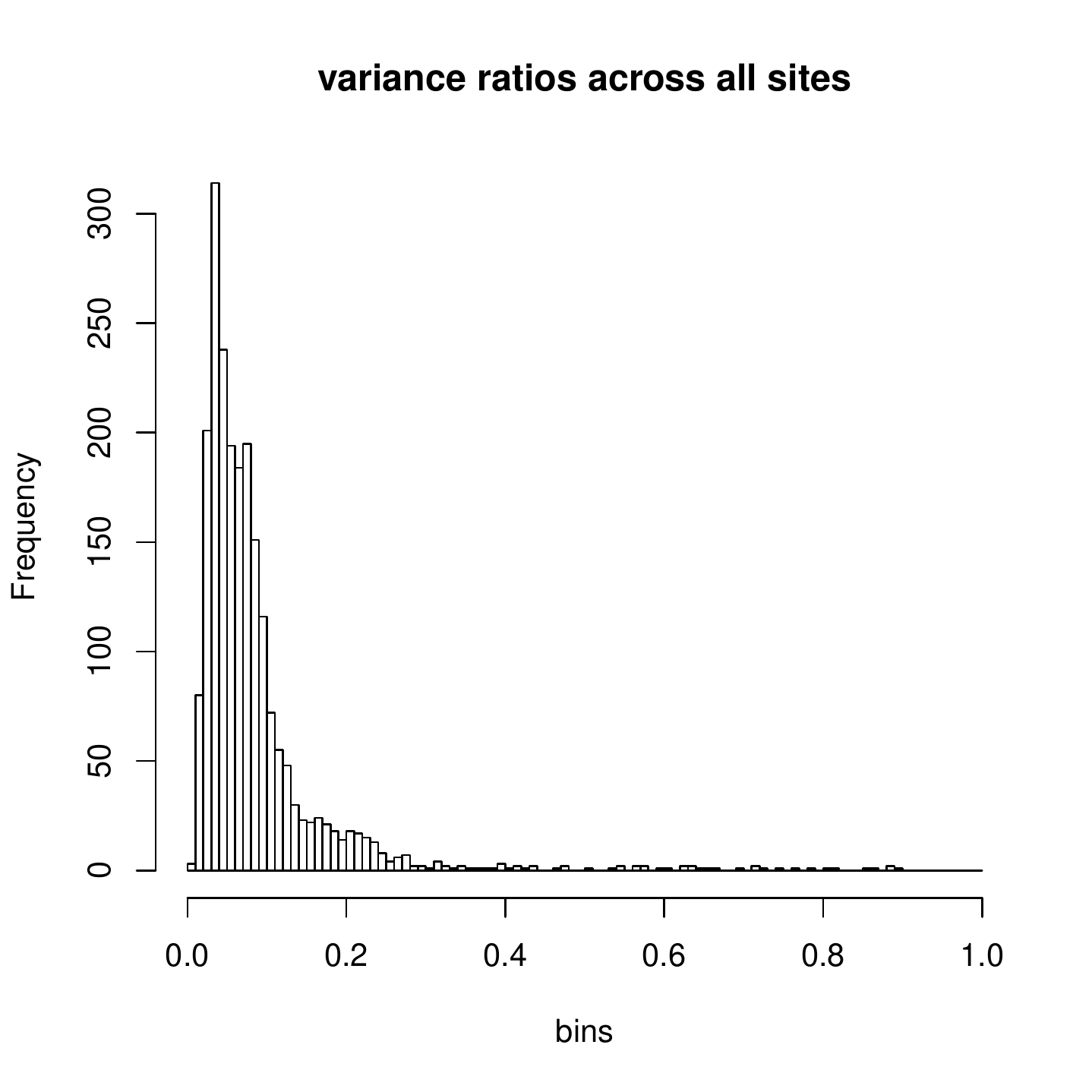}}  \scalebox{0.6}{\includegraphics{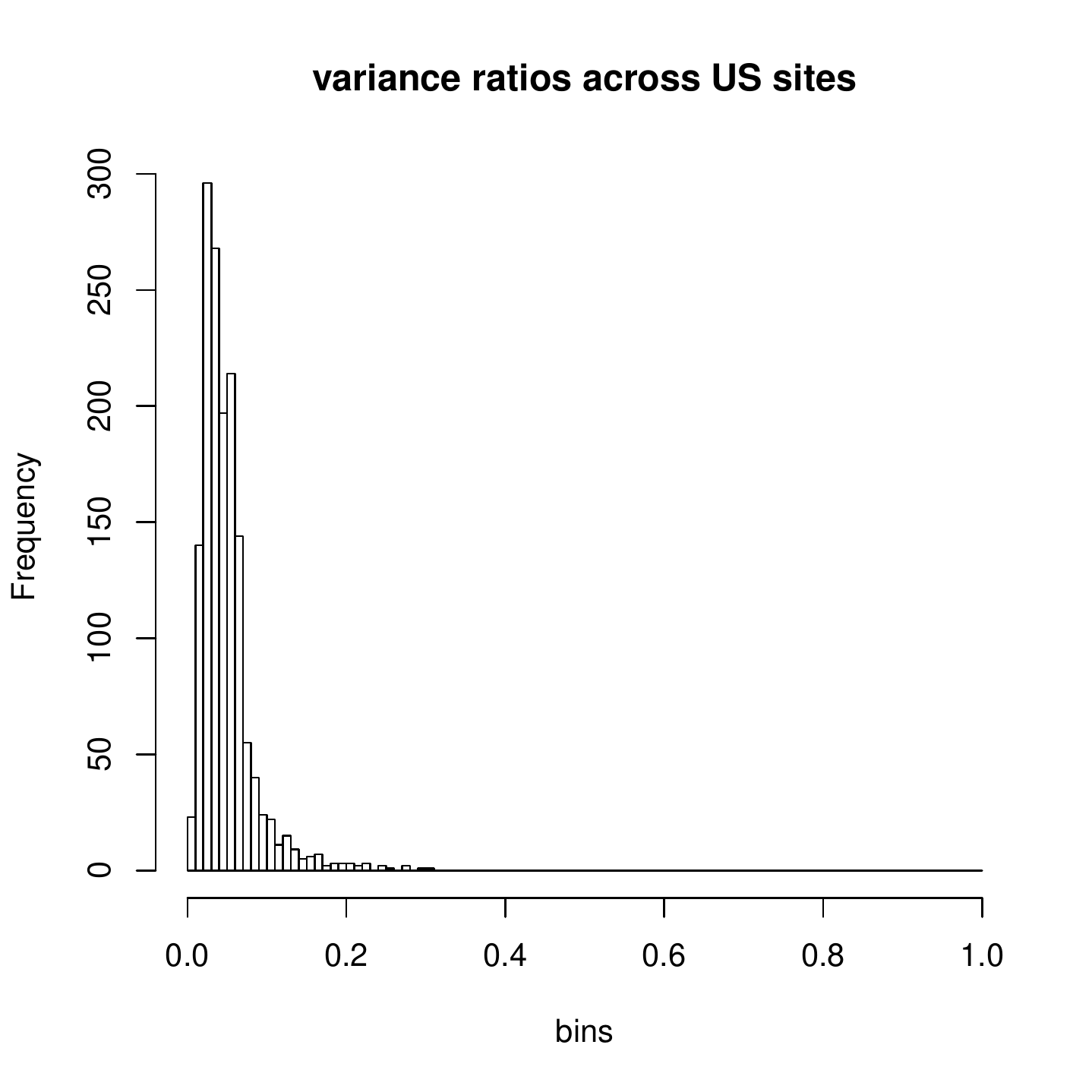}}}
 	\caption{Histograms of variance ratios for individual measures across all sites (left) and US sites (right), in bins of $0.01$.  }\label{fig:variance_ratio_histogram}
 \end{figure}

Figure \ref{fig:variance_ratio_histogram} shows a histogram of variance ratios.  The figure confirms these expectations, with a consistent distribution of variance ratios especially for US sites.  More detailed analysis (see Appendix A) showed that this variance ratio $q$ had a  $95\%$ confidence interval between $0.02$ and $0.1$ (for measures that don't depend on specific US cultural knowledge), between $0.02$ and $0.15$ (for all measures, but where replications are limited to US sites) and betweeen $0.02$ and around $0.35$ (for all measures and where replications take place in a diversity of cultural settings).  
Given the range of topics covered by the measures investigated here, we can conclude that a reasonable model for the variance ratio for exact replications would be one where $q$ falls in one or other of these ranges (depending on the cultural specificity of the experimental measures in question).

\section{Using distributional nulls }
\label{sec:joint_criterion}

The above results suggest that there is a consistent and reliable degree of random between-experiment variance in experimental means in these areas of social and cognitive psychology.    Suppose we have carried out a social psychology experiment, and obtained a result, $t_1$. What should we conclude?  Should we see this result as simply a consequence of random variation, or should we see this result as indicating a possible real effect? We can judge the likelihood that our result is simply a consequence of random variation by assessing it relative to a statistical model of such variation.   The point-form statistical model (which assumes within-experiment sampling variation around the mean $\mu$, but no variation in $\mu$ across experiments) is clearly not  appropriate in this situation, since we know that there is between-experiment variation in means.  Even if our result $t_1$ is very unlikely under the point-form null model, it may still be be a likely consequence of random variation in $\mu$.  

The distributional null model (which includes both within-experiment and between-experiment random variation) is more suitable in this situation: if we find that our result $t_1$ is unlikely to arise under a distributional null $q$ that appropriately includes both within-experiment and between-experiment random variation, we are justified in taking our result as suggesting a real effect.
In applying the distributional approach, however, we are faced with a problem:  we never know which value of $q$ is appropriate for a given experiment.  The best we can do is test against the range of values of $q$ that could reasonably hold.     In this section we describe a approach to assessing significance and replication across the range of possible $q$ values for a given experiment.

We assume that the experimenter sets a significance level $\alpha$ and a replication level $\beta > \alpha$  so that a result $t_1$ will be counted as real when its significance under the distributional null is less than $\alpha$ and when the probability getting a significant result in a repeat of that experiment (at the same level $\alpha$) is greater than $\beta$.  (Note that even when there is truly no effect, the probability of getting a significant result at level $\alpha$ in a repeat of our experiment is equal to $\alpha$; and so by assumption $\beta > \alpha$ necessarily holds.)   
From Equation \ref{eq:p_r},  the replication requirement $p_r(t_1,\nu) \geq \beta$ is met when
   \begin{equation*}
\begin{split}
\frac{\frac{qN}{1+qN }|t_1|-t_{crit}}{ \sqrt{\frac{1+2qN}{1+qN}}} \geq T_{\nu}^{-1}(\beta)
\end{split}
\end{equation*}
or, equivalently,  when  
\begin{equation*}
\begin{split} 
|t_1| \geq t_{rep}  =  \left(1+\frac{1}{qN}\right)\left( t_{crit}+ T_{\nu}^{-1}(\beta)\sqrt{\frac{1+2qN}{1+qN}} \right) \\
\end{split}
\end{equation*}
where $t_{rep}$ is the critical value for this replication criterion, just as $t_{crit}$ is the critical value for significance.  Take the maximum of these two critical values
\begin{equation*}
R_{q} = max(t_{rep},t_{crit})
\end{equation*}
and a result $t_1$ meets our replication and  significance criteria  $\beta$ and $\alpha$ when $|t_1| \geq R_{q}$.  $R_{q}$ thus represents the critical value for judging results as `real' in terms of both significance and replication (under distributional null $q$). 

Note that when $\beta = 0.5$, $T_{\nu}^{-1}(\beta) = 0$ and so $t_{rep} \geq t_{crit}$ for all $qN$, and we get
\begin{equation*} 
\begin{split}  
R_{q} = \left(1+\frac{1}{qN}\right) t_{crit} = \left(1+\frac{1}{qN}\right)T_{\nu}^{-1}(1-\alpha)\, \sqrt{1+qN}
\end{split}
\end{equation*}
Here the term
\begin{equation*}
\sqrt{1+qN}\left(1+\frac{1}{qN}\right)
\end{equation*}
is convex and unimodal with a minimum at $qN=2$.  This means that when $\beta = 0.5$ our critical value for judging results as `real' in terms of both significance and replication has a lower bound of
\begin{equation*}
R_q \geq   T_{\nu}^{-1}(1-\alpha)\frac{3\sqrt{3}}{2}
\end{equation*}
and so results $|t_1|$ less than this bound can never meet our significance and replication criteria for any distributional null. 	This suggests a straightforward rule of thumb for identifying `real' effects under distributional nulls: take the replication criterion $\beta =0.5$, and for a given significance level $\alpha$ any result where
\begin{equation*}
p > T_{\nu}\left(T_{\nu}^{-1}(1-\alpha)\frac{3\sqrt{3}}{2} \right)
\end{equation*} 
is rejected as not indicating a real effect.  For $\alpha = 0.05$, this bound is approximately $0.0005$ for $\nu = 10$, and
 $0.00005$ for $\nu = 40$; these results thus support the suggestion made by \citet{benjamin2019three} of replacing the $0.05$ threshold with a lower value; and indeed give a mathematical justification for such replacement (though the values obtained here are orders of magnitude lower than those suggested by Benjamin and Berger).

If a given result has a $p$ value less than this rule-of-thumb bound, we cannot conclude that this result meets our significance and replication criteria $\alpha$ and $\beta$.  To identify results that do meet these criteria for reasonable values of $q$, we note that, in fact, this critical value $R_q$ is convex and unimodal in $qN$ for all $\alpha$ and $\beta > \alpha$ (see Appendix B for details).  This means that for a given result $|t_1|$ if there exist values  $q_{1} <q_{2}$ such that
\begin{equation*}
|t_1| = R_{q_1} \textit {\,  and \, } |t_1| = R_{q_2}
\end{equation*}
then $|t_1| \geq R_q$ for all $q$ in the range $q_{1} \ldots q_{2}$: the result meets our replication and significance criteria for all $q$ in this range.  If such values exist, and if this range includes values of $q$ which seem reasonable given assumptions about cross-experiment variance in the experimental task, then we can conclude that our result meets significance and replication criteria $\alpha$, $\beta$ for reasonable values of cross-experiment variance. If such values do not exist, or fall outside the range of reasonable values for $q$, we can conclude that our result does not meet these criteria  $\alpha$, $\beta$.   Code to calculate this range for give $t$ values and $\alpha$,$\beta$ criteria is available online.\footnote{For review purposes, R code is included as a supplementary file.}

The value $\gamma= q_{2}$ here gives a rough measure of the generalisability of our result: if this value is large then our result is significant and replicable across a range of reasonable cross-experiment variance values, and we can say that the result is likely to hold even in experiments that differ to some degree from our original.  If this difference is small, then the result meets our significance and replication criteria only for experiments that match our original experiment quite closely.  Applying this distributional null approach to hypothesis testing, we get estimates for statistical significance (including within- and between-experiment variance, and not affected by issues to do with sample size) for probability of replication (not based on the assumption that the observed result $\overline{x_1}$ is equal to the true effect size) and for the degree of generalisation $\gamma$ of these results (based on observed rates of between-experiment variance).
 
 \section{Discussion}
 
 Our argument in this paper is for a move away from the standard point-form approach to NHST  to  a broader distributional-form NHST approach.  Standard point-form NHST suffers from various fundamental problems: the probability of getting a statistically significant result in the standard NHST  increases with sample size, irrespective of the presence or absence of a true effect; the standard NHST approach is overconfident in identifying effects as real (ignoring, as it does, the effects of random between-experiment variance); the standard NHST approach does not allow for correct estimates of the probability of replication of a given result.  A distributional-form approach to NHST avoids problems associated with  sample size,  allows for conservative rejection of the null, and allows meaningful and coherent estimation of the probability of replication. 
 
 We expect a number of objections to our argument.  The first objection concerns the use of a  distributional representation of the null hypothesis. `These distributional nulls are just  Bayesian priors in another form' we imagine the objection goes, `and Bayesian priors are subjective measures of belief, not objective probability estimates.  Subjective beliefs cannot enter into objective frequentist hypothesis testing'.  
 
 It is true that our distributional nulls have a mathematical form that is identical to a Bayesian prior.  It is not true, however, that these distributional nulls represent subjective measures of belief.  Instead, these distributional nulls play the same  role that point-form nulls play in NHST: for a given (distributional or point-form) null hypothesis $H$ we say `result $t$ would have a low probability of occurrence if null hypothesis $H$ were true, and so result $t$ is unlikely to be simply a consequence of random processes'.  Such  assertions do not require or reflect any subjective belief in the null hypothesis.  To put this response another way: since the logic of NHST is independent of the form of null hypothesis being used, our use of distributional rather than point nulls does not change the objective nature of such hypothesis testing.
 
 A second objection concerns statistical testing against distributional null hypotheses characterised by the parameter $q$ (representing  the between-experiment to within-experiment variance ratio).  `Researchers can adjust the distribution of this parameter  $q$ until they find a null hypothesis against which their observed results are statistically significant' we imagine the objection goes. `But this is simply a form of data-dredging or p-hacking: an attempt to find patterns in data that can be presented as statistically significant when in fact there is no real underlying effect'.  
 
 Our response here is to note that the use of distributional null hypotheses systematically reduces the occurrence of statistically significant results, relative to the point-form null.  This is because, as we saw earlier, the rejection region for a null hypothesis falls monotonically with the value of this ratio $q$: and so the maximum rejection region (and so the greatest chance of a statistically significant result) arises with the point-form null.  Given this, a better characterisation of this process of testing against nulls characterised by  this distributional parameter $q$ is one where we attempt to \textit{reduce} the chance of finding statistically significant results: where we are conservative in accepting patterns in data as being statistically significant, taking into account both between-experiment and within-experiment variance in our judgment.
 
 A third possible objection concerns the availability of alternative hypothesis-testing methods, such as the Bayes Factor test (commonly put forward as a replacement for NHST).  `While the NHST approach was appropriate in the last century', we imagine the objection goes, `today we have better statistical approaches to hypothesis testing; we don't need the NHST approach'. 
 
 We have two responses to this.   The first is to point out that the Bayes Factor approach, at least as it is typically used, is also based on the assumption of a point-form null hypothesis, and as such falls prey to many of the problems described above (problems to do with sample size, replication and so on).  The second and more general response is to point out that the NHST and Bayes Factor approaches ask two different questions.  The NHST approach involves a single null hypothesis  $H$ (the hypothesis that observed results are the chance consequence of some random process) and asks whether observed results are likely or unlikely under that null hypothesis.     The Bayes Factor approach, by contrast,  involves comparison of two contrasting hypotheses: Bayes Factor analysis asks whether experimental results give evidence for one hypothesis $H_0$ or for a specified alternative hypothesis $H_1$.   Both forms of question are useful and important: one, however, does not replace the other.
 
 \subsection{Related approaches}
 
 While the idea of using distributional rather than point-form null hypotheses will, we think, be relatively novel, it is worth pointing out that similar approaches are well known and commonly used in certain specific areas.  
 
 Perhaps the most important of these is that of `random effect meta-analysis'.  Meta-analysis attempts to assess the reality and size of some effect $d$ by systematically combining measures of that effect $d_i$ from different experimental studies.     A ‘fixed effect’ meta-analysis starts with the hypothesis that the true value of this effect $d$ is fixed at some point-form value, and that individual measures $d_i$ represent samples from a population with that fixed point-form value.  A `random-effect’ meta-analysis, by contrast, starts with the hypothesis that the effect $d$ has a distributional, rather than a point, form (this distribution is the random effect) and that individual measures $d_i$ represent samples from that distribution.   The idea that a distributional, rather than point-form, model of effects should be used in the medical and social sciences is well understood in the meta-analysis literature: to quote \citep{higgins2009re}
 \begin{quote}
 	Occasionally it may be reasonable to assume that a common
 	effect exists (e.g. for unflawed studies estimating the same physical constant). However, such an
 	assumption of homogeneity can seldom be made for studies in the biomedical and social sciences.
 	These studies are likely to have numerous differences, including the populations that are
 	addressed, the exposures or interventions under investigation and the outcomes that are examined. Unless there is a genuine lack of effect underlying every study, to assume the existence of
 	a common parameter [a fixed effect] would seem to be untenable.  
 \end{quote}
 There are, however, a number of major differences between the `distributional effect $d$' model used in random-effect meta-analysis and the `distributional null hypothesis' approach we describe.  Most obviously: our approach tests experimental results against a distributional null hypothesis, and does not involve any assumptions about the `true effect' $d$ underlying those results.  Random-effect meta-analysis, by contrast, makes no mention of the (point-form or distributional) null, and instead involves distributional assumptions about the effect size $d$, based on experimental results $d_i$.  
 In some ways this approach is similar to the statistical power model of replication described above, which begins with the assumption that the population effect size $d$ is equal to the observed result $d_1$ (disregarding the possibility that the observed result could have been produced under the null hypothesis).

 \section{Conclusions}
 
 Standard point-form NHST has a number of fundamental mathematical, experimental and practical problems.  These problems can be effectively and naturally addressed by moving to a distributional NHST approach.  The primary difference between the distributional and point-form NHST approach lies in the fact that the distributional approach accounts for random variation across experiments; point-form NHST, because it does not account for this type of random distributional variance, is overconfident in rejecting the null: results which arise purely as a consequence of random distributional variance  will be counted as statistically significant under point-form NHST. 
 
 The extent of this overconfidence depends on the degree of variance between experiments, which is a function of the number of confounding factors affecting experimental results.  Experiments in areas such as psychology, neuroscience, medicine, and so on (areas investigating very complex, interacting, and only partially understood systems) will necessarily be subject to many potential confounds, and so point-form judgements of statistical significance in these areas will be substantially and systematically overconfident: many results that are counted as real in point-form significance tests are in fact likely to have arisen as a consequence of random between-experiment variance.   
 
 A second problematic aspect of point-form NHST arises from the fact that the probability of getting a statistically significant result in the point-form NHST increases with sample size, irrespective of the presence or absence of a true effect.  This means that even in situations where confounding factors are very tightly controlled (and so distributional variance minimised) the point-form NHST approach will still be systematically overconfident, with results being judged statistically significant simply as a consequence of a large enough sample size, rather than a true effect.   This issue is particularly important in areas involving hypothesis testing across large data sets (again, medicine and to some extent psychology), where again results that are counted as real relative to standard significance criteria are in fact likely to have attained significance purely as a consequence of sample size.  
 
 Our analysis here shows how these problems can be addressed in a distributional NHST approach, at least for statistical tests based on the $t$ distribution under certain assumptions: normally distributed within-experiment data, normally distributed variance of the mean across replications, constant or relatively constant within-experiment variance in replications.

 \newcommand{\noop}[1]{}

 \pagebreak
 
 \section*{\centering Appendix A}
 
In this section we summarise the 9 experiments we use to estimate the variance ratio $q$, and the individual measures in each experiment  \citep[original sources and full presentation of these experiments are given in ]{klein2014investigating}. References to measures are in terms of the labels given in the replication dataset at \url{https://osf.io/wx7ck/}.  We also give results on variance ratio estimates for individual measures (grouped by topic) for all sites and US sites.  Results confirm those give in the main text. 

\subsubsection*{Experiment: Retrospective gamblers fallacy}
This study investigated whether the rarity of an independent, chance observation influenced beliefs about what occurred
before that event. Participants imagined that they saw a man rolling dice in a casino. In one condition, participants imagined witnessing three dice being rolled and all came up 6. In a second condition two came up 6 and one came up 3.  All participants then estimated, in an open-ended format, how many times the man had rolled the dice before they entered the room to watch him.   We computed the observed variance ratio $q$ for estimates in these two conditions (measures `gamblersfallacya': first condition and `gamblersfallacyb'; second condition).

\subsubsection*{Experiment: Sex differences in implicit math attitudes}
As a possible account for the sex gap in participation in science and math, this study asked whether women had more negative implicit attitudes toward math compared to arts than men did. Participants completed an Implicit Association Test (IAT) which measured associations of math and arts with positivity and negativity.  We  computed the observed variance ratio $q$ for the IAT association measure (measure `d\_art').  

\subsubsection*{Experiment: Relation between implicit and self-reported math attitudes}
In the same study, self-reported math attitudes were measured
with  'feeling thermometers'(preference ratings based on a 0–100
scale from cold/unfavorable to warm/favorable) assessing participants’
feelings of warmth toward math and arts as academic domains. We  computed the observed variance ratio $q$ for each of these measures (`mathwarm' and `artwarm').Participants also completed 6 semantic differential scales measuring
attitudes toward math, and 6 measuring attitudes towards arts, using dichotomous pairs of adjectives anchored
each end of a 7-point scale (good–bad, happy–sad, delightful– disgusting, beautiful– ugly, approach–avoid, and unafraid–
afraid).   We  computed the observed variance ratio $q$ for each of these $12$ `explicit IAT' questions (measures `iatexplicitart1' to `iatexplicitmath6').

\subsubsection*{Experiment: Sunk costs} 
Sunk costs are those that have already been incurred and cannot be recovered.  This study asked participants to imagine that they have tickets to see their favorite football team play an important game, but that it is freezing cold on the
day of the game.  Participants in group A were asked to imagine they had paid for the ticket, while those in group B were asked to imagine that the ticket had been free. Participants rated their likelihood of attending the game on a 9-point scale (1 = definitely stay at home, 9 = definitely go to the game).    We computed the observed variance ratio $q$ for the measure `sunkcosta' (likelihood rating in the first group) and   the measure `sunkcostb' (likelihood rating in the second group).

\subsubsection*{Experiment: Quote perception and attribution}
This study examined how an identical quote would be perceived if it was attributed to a liked or disliked individual. The quotation of interest was, ‘‘I hold it that a little rebellion, now and then, is a good thing, and as necessary in the political world as storms are in the physical world.’’ In one condition the quote was attributed  to  George Washington (liked individual); in the other to Osama Bin Laden (disliked individual).    We computed the observed variance ratio $q$ for answers on these two scales (measures `quotea': liked source, and `quoteb': disliked source).   We considered these measures to be culturally specific to a small degree, and so expected international replications to have slightly higher values of the variance ratio $q$ for international sites.

\subsubsection*{Experiment: Imagined contact with outgroups}
This study asked whether merely imagining contact with members of ethnic outgroups is sufficient to reduce prejudice toward those groups. In the study non-Muslim participants were assigned to either imagine interacting with a  Muslim stranger or to imagine that they were walking outdoors (control condition). Participants imagined the scene for one minute, and then described their thoughts for an additional minute before indicating their interest and willingness to interact with  Muslims in a four-item questionnaire, where response to each question fell on a 9-point scale.   For  replication in the predominately Muslim sample from Turkey the items were adapted so Christians were the outgroup target.   We  computed the observed variance ratio $q$ for responses to the each of the 4  questionnaire items (measures `imagined1',$\ldots$,`imagined4').   Again, we considered these measures to be possibly culturally specific, and so expected higher values of the variance ratio $q$ for international sites.

\subsubsection*{Experiment: Anchoring and estimation}
Anchoring occurs when participant estimates of some continuous value are influenced by a previously received `anchor' value.  This study involved 4 scenarios in which participants estimated size or distance after first receiving a number that was clearly too large (high anchor) or
too small (low anchor).    We computed the observed variance ratio $q$ for 8 measures in total (two for each scenario).   These scenarios asked about distance from San Francisco to New York City (measures `anchoring1a': small anchor value, and `anchoring1b': large anchor value ), population of Chicago  ( `anchoring2a': small anchor ,  `anchoring2b': large anchor), height of Mt. Everest  ( `anchoring3a': small anchor,  `anchoring3b': large anchor) and babies born per day in the US ( `anchoring4a': small anchor,  `anchoring4b': large anchor).  For replications in countries outside the US that use the metric system, anchors for cases `anchoring1a', `anchoring1b', `anchoring3a', and `anchoring3b' were converted to metric units and rounded.  Again, we considered these measures to be quite culturally specific (involving both US topics and requiring conversion to metric for sites outside the US), and so expected higher values of  $q$ for international sites.

\subsubsection*{Experiment: Flag priming and conservatism}
This study examined how subtle exposure to the American flag may increase conservatism among US participants. Participants were presented
with four photos and asked to estimate the time of day at which they were taken. In the flag-prime condition,
the American flag appeared in two of these photos. In the control condition, the same photos were
presented without flags.  4 distinct photos were used in each condition.  We computed the observed variance ratio $q$ for time estimates for photos with the flag present (measures `flagtimeestimate1',$\ldots$,`flagtimeestimate4') and those with the flag absent (measures `noflagtimeestimate1',$\ldots$,`noflagtimeestimate4').

Following time-estimation for these photos, participants completed an 8-item questionnaire assessing views toward various political issues (e.g., abortion, gun control, affirmative action) on a 7-point scale.   We computed the observed variance ratio $q$ for responses to these 8 questions (measures `flagdv1',$\ldots$,`flagdv8').

Three further questions  at the very end of the replication study tested possible moderators of these effects: (1) How much do you
identify with being American? (1 = not at all; 11 = very much), (2) To what extent do you think
the typical American is a Republican or Democrat? (1 = Democrat; 7 = Republican), (3) To what extent
do you think the typical American is conservative or liberal? (1 = Liberal; 7 = Conservative).   We  also computed the observed variance ratio $q$ for responses supplementary questions (measures `flagsupplement1',$\ldots$,`flagsupplement8').  We considered these measures to be culturally specific (involving the US flag and questions about US politics) and so expected higher values of the variance ratio $q$ for international sites.

\subsubsection*{Experiment: Currency priming}
This study investigated the extent to which merely exposing participants to money increases their endorsement of the current social system. Participants were first presented with demographic questions, with the background of the page manipulated between subjects. In one condition the background showed a faint picture of US \$100 bills; in the other condition the background was a blurred, unidentifiable version of the same picture. Next, participants completed an 8-question ‘‘system justification scale’’, with responses for each question falling on a 7-point scale.  For international replications  the US dollar was usually replaced with the relevant country’s currency, and the system justification questions were adapted to reflect the name of the relevant country.   We  computed the observed variance ratio $q$ for responses to the each of the 8 ‘‘system justification’’ questions (measures `sysjust1',$\ldots$,`sysjust8').  We considered these measures to be highly culturally specific, focusing as they do on the specific cultural `systems' in a given country, so expected higher values of  $q$ for international sites.

\subsection*{Results by measure}  Table \ref{tab:variance_ratio_all} shows the mean variance ratio across all sites, for measures grouped by topic into two sets: set 1 (less culture-specific measures) and set 2 (more culture specific).  For measures in set 1, the mean variance ratio was $0.05$ with little variation across groups, and $95\%$ of individual variance ratios fell between $0.02$ and $0.10$ (with variance ratios for measures in each group falling in similar ranges).  For measures in set 2, the mean variance ratio was higher ($0.14$) and the $95\%$ range was much wider (between $0.02$ and $0.64$).   Analysing variance ratios only for US sites (Table \ref{tab:variance_ratio_US}), we see the mean variance ratio across both sets of measures was $0.05$, with $95\%$ of all individual variance ratios falling between $0.01$ and $0.15$. These results confirm the general results seen in Figure \ref{fig:variance_ratio_histogram}.
 
\pagebreak
 
 \section*{\centering Appendix B}
 
 In this appendix we show that the critical value $R_q = max(t_{rep}, t_{crit})$ is convex and unimodal in $qN$.
 Rewriting our expression for $t_{rep}$ we get
 \begin{equation*} 
 \begin{split}  
 t_{rep} =T_{\nu}^{-1}(1-\alpha)\, \sqrt{1+qN}\left(1+\frac{1}{qN}\right)+ T_{\nu}^{-1}(\beta)\sqrt{\left(1+\frac{1}{qN}\right)\left(2+\frac{1}{qN}\right)} \\
 \end{split}
 \end{equation*}
 Here the term
 \begin{equation*}
 \sqrt{1+qN}\left(1+\frac{1}{qN}\right)
 \end{equation*}
 is convex and unimodal with a minimum at $qN=2$  and increasing monotonically as $qN$ moves away from that minimum, approaching infinity as $qN \rightarrow 0$ and $qN \rightarrow \infty$.    The term
 \begin{equation*}
 \begin{split}
 \sqrt{\left(1+\frac{1}{qN}\right)\left(2+\frac{1}{qN}\right)}\\
 \end{split}
 \end{equation*}
 declines with increasing $qN$, approaching infinity as $qN \rightarrow 0$ and approaching $\sqrt{2}$ as $qN \rightarrow \infty$.  Since by assumption $\beta > \alpha$, this means that $t_{rep}$ is also convex and unimodal, with a minimum at some value $ qN =m > 0$ and increasing monotonically as $qN$ moves away from $m$, approaching infinity as $qN \rightarrow 0$ and $qN \rightarrow \infty$.    Finally, from Equation \ref{eq:T_crit_q} we see that $t_{crit}$ increases monotonically with $qN$, from a minimum of $1-T_{\nu}(1-\alpha)$ at $qN=0$.  Together these points mean that $t_{rep} > t_{crit}$ holds for all $qN$ less than some transition point $u >0$, while $ t_{crit} > t_{rep}$ holds for values of $qN$ greater than this point $u$.  If this transition point $u$ is greater than or equal to $m$, then $R_q$ falls to its minimum at $qN = m$ and rises thereafter, and so is convex and unimodal in $qN$ with a minimum at $m$.   If that transition point is less than $m$, then then $R_q$ falls until $qN=u$, at which point $t_{crit} > t_{rep}$ holds and so $R_q$ rises thereafter.   $R_q$ is thus convex and unimodal in $qN$, as required.

 \begin{table}
 	\caption{Variance ratios across all replication sites for all grouped measures in set 1 (less culture-specfic) and set 2 (more culture-specific).  The variance ratio for each measure for each site is the ratio between between-site variance of means, and that sites within-site variance.  `Mean variance ratio' is the mean variance ratio across all sites for all measures in a group.  'datapoints' gives the number of variance ratios calculated (\#sites $\times$ \#measures within group).    } \label{tab:variance_ratio_all}
 	\centering
 	\begin{tabular}{p{4cm} rrrr}
 		\hline
 		& datapoints & mean variance ratio & 2.5\% quantile & 97.5\% quantile \\ 
 		\hline
 		gambersfallacy &  72 & 0.05 & 0.01 & 0.15 \\ 
 		d\_art &  36 & 0.05 & 0.04 & 0.07 \\ 
 		mathArtWarmth &  72 & 0.07 & 0.04 & 0.11 \\ 
 		iatexplicitart & 216 & 0.03 & 0.02 & 0.06 \\ 
 		iatexplicitmath & 216 & 0.07 & 0.04 & 0.10 \\ 
 		sunkcost &  72 & 0.05 & 0.02 & 0.10 \\ 
 		quotes &  72 & 0.06 & 0.04 & 0.10 \\ 
 		imagined & 144 & 0.07 & 0.04 & 0.12 \\ 
 		noflagtimes & 144 & 0.03 & 0.01 & 0.05 \\ 
 		flagtimes & 144 & 0.04 & 0.02 & 0.06 \\
 		\\
 		all set 1 & 1188 & 0.05 & 0.02 & 0.10 \\ 
 		\hline
 		\\
 		anchoring & 288 & 0.12 & 0.02 & 0.42 \\ 
 		flagdv & 288 & 0.10 & 0.03 & 0.26 \\ 
 		flagsupplement & 108 & 0.30 & 0.03 & 1.10 \\ 
 		sysjust & 288 & 0.14 & 0.06 & 0.26 \\ 
 		\\
 		all set 2 & 972 & 0.14 & 0.02 & 0.64 \\  
 		\hline
 		across both sets & 2160 & 0.09 &  0.02   &  0.35 \\
 		\hline
 	\end{tabular}
 \end{table}

 \begin{table}
 	\caption{Variance ratios across US replication sites for grouped measures, calculated as before.  } \label{tab:variance_ratio_US}
 	\centering
 	\begin{tabular}{p{3.5cm} rrrr}
 		\hline
 		& datapoints & mean\_variance\_ratio & lower\_quantile & upper\_quantile \\ 
 		\hline
 		gambersfallacy &  50 & 0.04 & 0.01 & 0.12 \\ 
 		d\_art &  25 & 0.05 & 0.03 & 0.07 \\ 
 		mathArtWarmth &  50 & 0.05 & 0.03 & 0.08 \\ 
 		iatexplicitart & 150 & 0.03 & 0.01 & 0.06 \\ 
 		iatexplicitmath & 150 & 0.05 & 0.04 & 0.07 \\ 
 		noflagtimes & 100 & 0.02 & 0.01 & 0.04 \\ 
 		sunkcost &  50 & 0.06 & 0.02 & 0.11 \\ 
 		quotes &  50 & 0.04 & 0.02 & 0.07 \\ 
 		imagined & 100 & 0.06 & 0.04 & 0.11 \\ 
 		flagtimes & 100 & 0.03 & 0.01 & 0.06 \\ 
 		anchoring & 200 & 0.05 & 0.02 & 0.22 \\ 
 		flagdv & 200 & 0.08 & 0.02 & 0.22 \\ 
 		flagsupplement &  75 & 0.04 & 0.01 & 0.10 \\ 
 		sysjust & 200 & 0.05 & 0.02 & 0.08 \\ 
 		\hline
 		all & 1500 & 0.05 & 0.02 & 0.15 \\  
 		\hline
 	\end{tabular}
 \end{table}

    \end{document}